# Trajectory-Based Urban Air Mobility (UAM) Operations Simulator (TUS)

Euclides C. Pinto Neto, Derick M. Baum,
Jorge Rady de Almeida Junior, João Batista Camargo Junior, Paulo Sérgio Cugnasca*
Safety Analysis Group
Polytechnic School of University of São Paulo
Brazil

## Abstract

Nowadays, the demand for optimized services in urban environments to provide better society wellness is increasing. In this sense, ground transportation in dense urban environments has been facing challenges for many years (e.g., congestion and resilience). One import outcome of the effort made toward the creation of new concepts for enhancing urban transportation is the Urban Air Mobility (UAM) concept. UAM aims at enhancing the city transportation services using manned and unmanned vehicles. However, these operations bring many challenges to be faced, e.g., the interaction between the controller agent and autonomous vehicles. Furthermore, trajectory planning is not a simple task due to several factors. Firstly, the trajectories must consider a reduced minimum separation as eVTOL vehicle are expected to operate in complex urban environments. This leads the trajectory planning process to observe safety primitives more restrictively once the airspace is expected to comport many vehicles that follow small minimum separation standards. Thereupon, the main goal of the Trajectory-Based UAM Operations Simulator (TUS) is to simulate the Trajectory-Based UAM operations in urban environments considering the presence of both manned and unmanned eVTOL vehicles. For this, a Discrete Event Simulation (DES) approach is adopted, which considers an input (i.e., the eVTOL vehicles, their origin and destination, and their respective trajectories) and produces an output (which describes if the trajectories are safe and the elapsed operation time). The main contribution of this simulation tool is to provide a simulated environment for testing and measuring the effectiveness (e.g., flight duration) of trajectories planned for eVTOL vehicles.

## 1 Introduction

Nowadays, the demand for optimized services in urban environments to provide better society wellness is increasing. In this context, Smart City is a concept with different interpretations [1] and presents a fundamental relationship with Internet of Things (IoT). Conversely, one possible definition of Smart City is presented in [2]: "connecting the physical infrastructure, the IT infrastructure, the social infrastructure, and the business infrastructure to leverage the collective intelligence of the city". This relatively new concept brings the idea of using interconnected smart devices to optimize several services in urban environments [3] [4] [5]. Smart Cities incorporate different service sectors, such as Smart Governance, Smart Mobility, Smart Utilities, Smart Buildings, and Smart Environment [6]. Furthermore, one essential service sector for society nowadays is transportation.

Ground transportation in dense urban environments has been facing challenges for many years (e.g., congestion and resilience). In the past few years, the industry and the scientific communities have invested resources towards the creation of new ideas to improve the urban transportation performance. One import outcome of this process is the conception of smart air transportation system, which composes the Urban Air Mobility (UAM) concept [7]. UAM is a transportation concept in which people and goods are moved around metropolitan areas in air vehicles in a diverse airspace environment [8]. There are projects under development and in operation nowadays into the industry in the context of air transportation services in Smart Cities and IoT, such as the Uber Elevate [9], the Amazon Prime Air [10], and NASA's Unmanned Traffic Management (UTM) [11]. These projects aim at enhancing the city transportation services using manned and unmanned vehicles. However, there is a concern regarding the usage of such technologies from the safety perspective. Furthermore, although using autonomous vehicles brings several benefits to society, such as saving lives, increased mobility, reducing the cost of congestion, reducing energy use and fuel emissions and improving land use (for ground vehicles) [12], the acceptance of these technologies from society is a challenge nowadays from different perspectives.

---

*euclidescpn@usp.br, derick.baum@usp.br, jorgerady@usp.br, joaocamargo@usp.br, cugnasca@usp.br

Moreover, air transportation is vital for society nowadays and it is increasing gradually [13]. Although this growth leads to a higher revenue, more complex environments are faced. Indeed, there are many challenges to be faced by authorities in the following years, especially regarding airspace safety and efficiency. Furthermore, new technologies and concepts are under development nowadays for improving the airspace operation. For instance, the Unmanned Aircraft Systems (UAS) [14] and Decision Support Tools (DST) for Air Traffic Controllers (e.g. Arrival and Departure managers) [15]. These new technologies and concepts are beneficial from many perspectives, such as safety, efficiency and airspace capacity.

One concept brought from the conventional air transportation system to Urban Air Mobility (UAM) is the centralized control for regions of the airspace. In this paradigm, agents (human or not) are responsible for ensuring the air traffic to move safely and efficiently throughout the airspace with the support of Decision Support Tools (DST) (e.g., arrival managers). However, problems may arise when the operator deals with dense regions (i.e., regions with an elevated number of aircraft) and with uncertainty (e.g., emergency situations). Thus, another source of problems relies on the interaction between the controller agent and autonomous vehicles, i.e., the interaction between the human and machine, and, thus, in the interaction between autonomous systems [16] [17]. In fact, during the early stages of the usage of autonomous vehicles (e.g., Unmanned Aircraft Systems), the sparsity between fully autonomous machines is a key safety feature once there is an understanding nowadays that manned vehicles are more predictable [18]. The adoption of this restriction may lead (i) to a gradual acceptance of those technologies from the society; (ii) a more controllable environment for authorities and regulators; (iii) a reduced impact on Air Traffic Controllers (ATCo); and (iv) a more resilient air traffic distribution once failures in autonomous systems can be treated separately.

Moreover, the transportation of people in urban airspace is not a new concept since helicopters were used in such operations [19]. This experience in air transportation in urban environments brought many innovations to the helicopters [20] and is valuable for successful UAM operations. However, UAM is expected to comport a much larger number of vehicles, which inevitably brings new challenges for safe and efficient operations. The potential of transforming the way transportation by offering redundant connections between different regions highlights viable approaches for dealing with the traffic congestion problems [21].

Several projects are being conducted for implementing the UAM concept. For instance, Uber Elevate aims to implement an Urban Air Mobility (UAM) ridesharing network in cities across the world [9]. In order to accomplish this, Uber has defined a set of requirements for Electric Vertical Takeoff and Landing (eVTOL) aircraft (e.g., size, speed, cruise condition, energy storage, and operation range). Such aircraft is considerably less expensive in comparison to helicopters [22], and is based on advances in electrical and distributed propulsion and the increase in energy densities of batteries [23]. Thus, this paradigm is becoming more affordable and accessible to society [9]. Moreover, the UAM initiative enables different missions to be conducted in an urban environment [24], such as on-demand air-taxi, air cargo, and Emergency medical evacuations. Furthermore, these missions involve delivering people and goods from different points of the urban environment and demand a proper trajectory planning process in order to be safe and efficient.

Furthermore, the problem of congestion in urban environments will become more significant with the growth of populations and urbanization [25]. There is a lack of physical space to build new transportation infrastructure (e.g., bridges). However, the integration of UAM aircraft must be conducted with minimal impact on existing operations. For a safe and viable integration, a set of principles are stated by NASA in the development of UAM airspace integration concepts, technologies, and procedures [24] [26] [19]. For instance, the compliance with regulatory requirements for vehicle-level and system-level safety and security and the inclusion of minimal additional requirements on existing airspace.

Thereupon, NASA proposed the UAM Maturity Levels (UML) for measuring the evolution of UAM [27]. Each level represents a state of UAM operation, which vary from an initial to mature and the challenges of each phase are related to the vehicles (e.g., autonomy), the airspace (e.g., air traffic management), and community (e.g., social acceptance).

This research considers early stages of UAM operation (e.g., in terms of aircraft separation rules), but is intended to consider a dense airspace.

Several companies are working in the development of UAM vehicles and in UAM operational concepts. Indeed, this is endorsed by the possibility of considerably reducing the total operating costs in comparison to many helicopters used nowadays [28] [29]. Besides, there are other benefits offered by UAM operations for society, such as the reduction of trip time [24] [9], more predictable trip duration [19], and an increase into the total transportation system capacity once UAM is expected to operate in parallel with ground-based systems.

However, trajectory planning is not a simple task due to several factors. Firstly, the trajectories must consider a reduced minimum separation as eVTOL vehicle are expected to operate in complex urban environments. This leads the trajectory planning process to observe safety primitives more restrictively once the airspace is expected to comport many vehicles that follow small minimum separation standards.

Secondly, dense environments tend to be expected in UAM operations. This harden the process of planning trajectories once many deviations from the so called direct flight tend to be required. These deviation in the optimal trajectory may be difficult to be found since the elevated number of vehicles (and their required minimum separation) reduces the available regions.

Thereupon, the process of measuring trajectories in terms of safety and efficiency is not simple. In fact, dealing with multiple trajectories is challenging once conflicts can be generated in different areas. Furthermore, predicting future positions of vehicles in UAM operations is not simple because a extensive investigation on their flying dynamics must be conducted.

Thirdly, the trajectory planning process must be conducted at real-time. Indeed, this process is intended to be conducted strategically, i.e., before the flight starts. Conversely, there are situations in which tactical interventions in trajectory planning are required (e.g., emergencies). In those cases, the process of planning trajectory an measuring its effectiveness must last a short period of time in order to maintain the operations safe.

Furthermore, the inclusion of Unmanned Aircraft Systems (UAS) into UAM operations is a challenge for many reason (e.g., social acceptance). In order to include these aircraft as self-piloted eVTOL vehicles, alternative approaches must be considered (e.g., increased longitudinal separation), which may also harden the process of planning trajectories for other aircraft.

The main goal of the Trajectory-Based UAM Operations Simulator (TUS) is to simulate the Trajectory-Based UAM operations in urban environments considering the presence of both manned and unmanned eVTOL vehicles. For this, a Discrete Event Simulation (DES) approach is adopted, which considers an input (i.e., the eVTOL vehicles, their origin and destination, and their respective trajectories) and produces an output (which describes if the trajectories are safe and the elapsed operation time). The main contribution of this simulation tool is to provide a simulated environment for testing and measuring the effectiveness (e.g., flight duration) of trajectories planned for eVTOL vehicles.

The main goal of this research is to propose a simulation framework for measuring safety and effectiveness of Trajectory-Based UAM operations in urban environments considering the presence of both manned and unmanned eVTOL vehicles. This simulation tool is called Trajectory-Based UAM Operations Simulator (TUS). Thereupon, a Discrete Event Simulation (DES) approach is adopted, which considers an input (i.e., the eVTOL vehicles, their origin and destination, and their respective trajectories) and produces an output (which describes if the trajectories are safe and the elapsed operation time). The main contribution of this simulation tool is to provide a simulated environment for testing and measuring the effectiveness (e.g., flight duration) of trajectories planned for eVTOL vehicles.

This paper is organized as follows: Firstly, Section 2 presents the related works. Then, Section 3 presents the definitions and characteristics of Urban Air Mobility (UAM). After that, Section 4 presents the main contribution of this research, the Trajectory-Based UAM Operations Simulator (TUS). Finally, Sections 5 and 8 presents the experiments and conclusions of this research, respectively.

## 2 Related Works

This section presents some works that are related to the contribution presented in this paper. It is worth noting that the contribution of this research is intended to be complementary to the contributions presented in the literature, i.e., the related works present outstanding contributions that are used as the foundation for our proposal.

The authors in [30] aim to make ATM research results more comparable by sharing tools and data using a fully open-source and open-data approach to air traffic simulation. The main challenges were to achieve a high fidelity (e.g., aircraft performance) and to increase the adoption by the community by keeping the program as simple as possible.

Consideration an adoption of this platform by many users, this can be considered a useful tool concerning innovation and applications development (e.g., providing specific methods for different problems). The paper describes the difficulties faced when using a fully open-data and open-source policy in this area.

Tra et al. [31] present conflict rate models to determine the intrinsic safety of airspace designs that consider conflicts between aircraft in different flight phases. Fast-time simulations were performed for several different layered airspace concepts considering unstructured airspaces. The results indicate that the models can estimate the conflict rate for high traffic densities. When comparing the different layered airspace concepts tested, the model predicted, and the simulation results, a clear safety improvement when the heading range is reduced can be perceived. Thus the models can be used to study the effect of airspace design parameters on the safety of airspace concepts.

In [32], numerical simulations are used in order to demonstrate the effectiveness of the proposed conflict management approach, which ensures conflict avoidance among aircraft and transition of aircraft into adjacent airspace. Complexity is also modeled as aircraft heading and speed deviations in a given sector to avoid conflicts. Thus, a specific architecture is proposed for planning aiming to minimize complexity for the neighbor sectors. More specifically, the conflict avoidance problem can be seen as a mixed integer Linear Programming (LP) subject to maneuver constraints. Thus, the aircraft can find the optimal solution by solving the LP problem and resolve conflicts among the aircraft and reduce the air traffic complexity of the neighbor sectors. Moreover, the proposed conflict management algorithm can identify the optimal conflict resolution maneuver of aircraft in near real-time considering multi-sector environments. The authors intend to investigate the relationship between the maneuver constraints and traffic complexity in future works.

Borener et al. [33] present Unmanned Aircraft Systems (UAS) modeling and simulation that consider a use case scenario that is consistent with the FAA's concept of operations for integration of UAS into shared airspace and employs sensing (using actual radar track traffic data) and medium fixed wing UAS. The proposed simulations offer functionality related to UAS operations (e.g., 'detect and avoid', mission profiles, positional variance, performance variance, fuzzy conflicts, variation in time spent in communication, and deviation from planned or intent profiles). The simulations conducted were based on the RAMS plus fast-time simulator tool and aimed to evaluate the separation indices and the number and severity of the separation events. The experiments were conducted in a simulated Houston Metroplex Environment. The results obtained showed that multiple UAS would considerably increase the likelihood of separation event and separation critically indices, and the usage of the "return to departure land site" contingency operation in case of failures in UAS communication link has a considerable impact on separation events. A difficulty faced by researchers, though, is the lack of historical data of UAS operation.

The authors in [34] focus on simulation-based Air Traffic Controller (ATCo) training using the Beginning to End for Simulation and Training (BEST). This proposal is a simulation tool adopted in training organizations in Europe. Although BEST simulator covers all levels and types of training (e.g., basic, validation, and conversation refresher), this research is focused on the basic part of the initial training. Furthermore, insights into challenges the candidates face when mastering the techniques of performance-based training are presented. ATCos are responsible for guiding aircraft through the airspace and their extensive training process considers practical exercises performed on computer devices. This process is divided into three phases: the initial training (basic and rating training), unit training (transitional, pre-on-the-job and on-the-job training) and continuation training (conversion and refresher training). Moreover, BEST simulator meets all the objectives and requirements prescribed for the provision of basic ATCo training.

# 3 Urban Air Mobility (UAM)

In this Chapter, we present the UAM concepts as well as the characteristics related to its operation. Firstly, the definition of Urban Air Mobility (UAM) is presented. Then, the operational constraints considered in these scenarios are illustrated. After that, the challenges faced nowadays in UAM context are presented. Finally, the final remarks of this chapter are depicted.

## 3.1 Urban Air Mobility (UAM) Definition

In this section, a discussion on Urban Air Mobility (UAM) definition is conducted. Firstly, the concept of this new paradigm is presented. Then, the historical background that can be beneficial for successful UAM operations is presented. After that, the principles and the UAM Maturity Levels (UML) are described. Finally, a discussion on the benefits of UAM for the transportation system is conducted.

### 3.1.1 Concept

Urban air mobility (UAM) is a transportation concept in which people and goods are moved around metropolitan areas in air vehicles in a diverse airspace environment [8]. This concept can be seen as on-demand, highly automated, passenger or cargo-carrying air transportation services within an urban environment [29].

In this context, different constraints are taken into account (e.g., weather conditions and restricted areas). The operations considered herein presents some similarities with helicopter operations (e.g., goals, performance, and altitude of operation). However, UAM consists of much larger numbers of aircraft transporting considerably more people in a wider transportation system [8].

This new paradigm of operation employs quieter and efficient manned and unmanned vehicles to conduct on-demand and scheduled operations. Examples of operations are emergency medical evacuations, rescue operations, humanitarian missions, and weather monitoring [24]. Furthermore, another outstanding example of UAM operation relies on passenger transportation, which represents significant time savings in many cases [19].

Moreover, there is an effort being made by different stakeholders toward developing the UAM concept. For instance, Uber published a white paper describing their vision for an air taxi service in its Uber Elevate initiative [19] [9]. Therefore, there are other companies working on technologies for air-taxi [35] [36] [37] [38].

### 3.1.2 Past Operations and Possible Inspirations for UAM

The transportation of people in urban airspace is not a new concept. Back to 1947, Los Angeles Airways used helicopters to transport people and mail in Los Angeles [19]. This broad experience in air transportation in urban environments brought many innovations to the helicopters as well as to the way the operations are conducted (e.g., air traffic management and integration challenges) [20].

This experience in urban scenarios is valuable for successful UAM operations. Indeed, this new paradigm is intended to comport a much larger number of vehicles, which inevitably brings new challenges for safe and efficient operations. On the other hand, there is a potential of transforming the way that transportation is conducted in cities by including a new operational layer. This alternative option offers solid connections between different regions as a viable approach for dealing with the traffic congestion problem currently faced [21].

Besides, the UAM initiative expects Electric Vertical Take-off and Landing (eVTOL) vehicles to operate extensively. These vehicles are considerably less expensive in comparison to helicopters [22]. This new technology is based on advances in electrical and distributed propulsion as well as an increase in energy densities of batteries [23]. In this sense, this new paradigm is becoming more and more cost-effective and accessible to society, and it is expected to be less expansive than car rides in some cases [9]. Finally, there is a new trend towards making short air transportation more affordable to the general public [29].

### 3.1.3 Principles

In the next decades, the problem of congestion in urban environments will become more significant with the growth of populations and urbanization [25]. In many regions, there is a lack of physical space to build new transportation infrastructure. Thus, UAM stands out as a solution for mitigating this problem.

However, the integration of UAM eVTOL vehicles (and technologies) must be conducted safely and efficiently, with minimal impact on existing operations. In this sense, this research assumes a set of principles followed by NASA in the development of UAM airspace integration concepts, technologies, and procedures [24] [26] [19]. These principles are:

1. "UAM should require minimal additional ATC infrastructure (e.g., radar systems, controller positions) and minimal changes to FAA automation systems used for ATC";

2. "UAM should impose minimal additional workload on controllers beyond their current duties for existing airspace users";

3. "UAM should impose minimal additional requirements or burdens on existing airspace users beyond equitable access to airspace resources";

4. "UAM will meet the regulatory requirements for vehicle-level and system-level safety and security, such as timely and assured data exchange and the elimination of single points of failure and common failure triggers";

5. "UAM will be resilient to a wide range of disruptions, from weather and localized sub-system failures (e.g., a single vehicle or software tool) to widespread disruptions (e.g., GPS failure)";

6. "UAM will economically scale to high-demand operations with minimal fixed costs";

Table 1: NASA's UAM Maturity Levels (UML).

| State | UML | Description |
|---|---|---|
| Initial | 1 | Early Operation Exploration and Demonstrations in Limited Environments |
| | 2 | Low Density and Complexity Commercial Operations with Assistive Automation |
| Intermediate | 3 | Low Density, Medium Complexity Operations with Comprehensive Safety Assurance Automation |
| | 4 | Medium Density and Complexity Operations with Collaborative Responsible Automated Systems |
| Mature | 5 | High Density and Complexity Operations with Highly-Integrated Automated Networks |
| | 6 | Ubiquitous UAM Operations with System-Wide Automated Optimization |

Source: [27].

7. "UAM will support user flexibility and decision making to the greatest extent possible and enforce airspace structure and prescriptive procedures only as necessary to meet the above principles";

Thereupon, these principles lead the operation to be feasible and compatible with the current transportation system.

### 3.1.4 UAM Maturity Level (UML)

In order to measure the evolution of UAM, NASA proposed the UAM Maturity Levels (UML) [27], illustrated in Table 1. Each level represents a state of UAM operation, which vary from an initial to mature. The challenges of each phase are related to the vehicles (e.g., autonomy), the airspace (e.g., air traffic management), and community (e.g., social acceptance).

Initial State

UML 1 represents the early operational exploration and demonstrations in limited environments. This represents the first stage of UAM integration into the airspace and considers a very controlled environment. The focus in this phase is to evaluate the aircraft from an operational perspective, to define and conduct certification processes, to employ traditional procedures of the airspace and to understand the perception of the community regarding this new paradigm.

UML 2 represents low density and complexity commercial operations with assistive automation. In this phase, the aircraft are expected to have a certification process, UAM corridors through controlled airspace are expected to be built, and operations with small UAM networks serving urban periphery are expected to be conducted.

Intermediate State

UML 3 represents low density with medium complexity operation with comprehensive safety assurance automation. In this phase, the operations are conducted into urban centers. Thus, automation for scalable and weather-tolerant operations takes place. Finally, local regulations are also expected to be considered.

After that, UML 4 presents a medium density and complexity operations with collaborative and responsible automated systems. Simultaneous operations are expected to be conducted, as well as expanded networks including high-capacity UAM ports. Finally, low-visibility operations are also expected to be conducted.

Mature State

In UML 5, high density and complexity operations with highly-integrated automated networks are considered. In this case, many simultaneous operations are expected to be conducted. Thus, large-scale, highly-distributed networks are considered. Furthermore, autonomous aircraft are expected to be part of the fleet.

Finally, UML 6 presents ubiquitous UAM operations with system-wide automated optimization, in which a substantial amount of simultaneous operations are conducted, and the UAM has a high social acceptance.

### 3.1.5 Transportation Benefits

Several companies have placed a considerable amount of resources in the development of UAM aircraft and operational concepts [29]. For instance, in some cases, the cost of UAM services may be comparable to current automobile-based ridesharing services [9].

Furthermore, self-piloted electric VTOL aircraft can reduce the total operating costs considerably in comparison to many helicopters used nowadays. This is due to (i) a reduction in fuel and maintenance costs associated with the electric propulsion technology; (ii) no costs for onboard piloting or additional equipment; and (iii) cost reductions enabled from advanced manufacturing processes [28] [29]. Porsche [22] states that compared to a traditional helicopter, eVTOL aircraft are quieter (4x), more reliable (15x), safer (2x), and less expensive (10x).

In terms of travel time, the same report compares the trip duration of cars and eVTOL vehicles. For example, consider a 40km trip. Cars can reach closer regions more quickly, once eVTOL vehicles have deal with first mile trips (which takes the user from origin - e.g., home - to the closest skyport) and boarding duration before starting the actual flight. However, in the range of 10km to 25km, the fastest option will depend on the characteristics of the region. Thereupon, we can infer that trips with more than 25km of extension tend to present a smaller duration when performed with eVTOL vehicles. Furthermore, this report also highlights that a reasonable duration of a 40km trip conducted by a car is 45 minutes, whereas eVTOL vehicles would perform it in 26 minutes. Finally, the benefits regarding time-saving refers to more extensive trips.

Moreover, there are also potential benefits to the overall transportation system. "The most fundamental motivation for UAM is transporting people and goods around metropolitan areas at door-to-door speeds greater than and in a manner complementary to current air and ground transportation capabilities" [24].

UAM has the potential to conduct trips with half (or less) the duration of ground transportation trips [24] [9]. The potential time savings of UAM could represent a large number of minutes, considering multiple users and an effective Air Traffic Management (ATM) system for UAM.

Another benefit relies on the reduction of uncertainty in travel duration relative to ground transportation in highly-congested cities [19]. Furthermore, the total transportation system capacity is expected to increase since UAM is expected to operate in parallel with ground transportation. Similarly, another benefit is that UAM tends to require smaller ground infrastructure. Instead of building large-scale ground connections (e.g., roads, overpasses, and bridges), UAM demands vertiports that are significantly smaller.

Indeed, from the infrastructural perspective, UAM can be implemented more quickly and at a lower cost compared to other ground-based models. Conversely, UAM demands significant electrical infrastructure for power generation once many operations are expected to be conducted by electric-powered vehicles[2] [9] [24] [19].

Finally, traditional on-demand air transportation services (such as helicopter services) have been used at high costs. However, there is a trend toward making short air trips widely affordable and available to the general public [29].

## 3.2 Operational Constraints

UAM operations present specific characteristics in terms of constraints. In this sense, we present the operational constraints considered in these operations. Firstly, the aircraft scope is depicted. Finally, the mission scope is highlighted, and the hazard scope is described.

### 3.2.1 Aircraft Scope

Several initiatives are under development for implementing the UAM concept in real-world environments. Among these initiatives, Uber Elevate plays an important role in this market. This Uber project aims to implement an Urban Air Mobility (UAM) ridesharing network in cities across the world [9]. In order to attend this demand, Uber has defined a set of requirements for Electric Vertical Takeoff and Landing (eVTOL) vehicles based on the analysis of current and predicted demand conducted. The main goal of this analysis was to understand the capability of the technology and to create an optimized user experience [39]. Thus, the aircraft scope considered in this research follows the eVTOL vehicle requirements proposed by Uber since it represents a realistic and viable approach for UAM operations.

The vehicles requirements considered in this research are [39]:

- · VTOL: Vehicles considered for this operation must present vertical takeoff and vertical landing capabilities with short hover duration;

---

[2]Note that this challenge may also be faced by electric-powered cars in ground-based transportation systems.

· Size: Vehicle must not exceed 45' in its max dimension, and the max takeoff weight must not exceed 7,000 lbs. Note that this definition highlights that small vehicles (e.g., package delivery drones) are not considered in our scope;

· Range: Vehicles should be able to fly 60 miles (or 52.14 Nautical Miles) and store enough energy to fly the reserve mission;

· Energy Storage: All aircraft considered herein are capable of storing energy for operations;

· Cruise Condition: Vehicles should operate in a cruise speed between 150mph and 200mph at 1000ft Above Ground Level (AGL) [9] [40].

· Ground Taxi: Vehicles must be able to perform a ground taxi without rotors spinning;

· Safety: Vehicles must be able to perform a safe vertical landing in emergencies;

· Skyport: Vehicles on the network will operate only from Skyports[3];

· Charging: the vehicles considered must be able to interface with rapid charging infrastructures and are expected to charge for 5 to 15 minutes between flights;

· Pilots: Uber is expecting that "vehicles will have the avionics and sensors needed for autonomous flight" [39]. However, the aircraft will be piloted in the first years while data is collected to enhance and prove the safety of autonomous systems. Furthermore, we consider two types autonomous eVTOL vehicles in this research: self-piloted eVTOL vehicles (also referred as Unmanned Aircraft Systems -UAS), which is represented by a fully autonomous eVTOL vehicle, and Remotely Piloted Aircraft System (RPAS), which represents an eVTOL vehicle piloted by human operators in a ground station;

· Network Communications: the vehicle must be able to receive data (e.g., flight plans) as well as to send current state (e.g., position and battery);

### 3.2.2 Mission Scope

The missions considered in this research follows the requirements proposed by Uber Elevate [39]. Figure 1 illustrates the segments of an eVTOL mission in this context. For each segment, Table 2 depicts its horizontal speed and Above Ground Level (AGL) ending altitude.

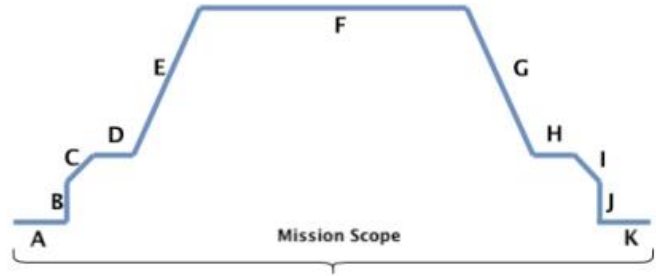

Figure 1: Mission scope of this research.

Source: Adapted from [39].

Moreover, the UAM initiative enables different missions to be conducted in an urban environment. A partial list of missions is presented in [24] as follows:

· "On-demand air taxi operations moving people between fixed or ad hoc locations";

· "Air cargo operations moving goods between warehouses and stores";

· "Regularly scheduled 'air metro' operations transporting passengers between a set of fixed locations";

· "Emergency medical evacuations, rescue operations, and humanitarian missions";

· "Weather monitoring";

· "Ground traffic assessment".

---

[3]Skyports are the airports of eVTOL vehicles. Note that these stations also offer charging infrastructure.

Table 2: Performance of the aircraft in each segment.

|   | Segment | Horizontal Speed (mph) | AGL Ending Altitude (ft) |
|---|---|---|---|
| A | Ground Taxi | 3 | 0 |
| B | Hover Climb | 0 | 40 |
| C | Transition + Climb | 0 to 1.2*Vstall | 300 |
| D | Departure Terminal Procedures | 1.2*Vstall | 300 |
| E | Accel + Climb | 1.2*Vstall to 150 | 1000 |
| F | Cruise | 150 to 200 | 1000 |
| G | Deceleration + Descend | 150 to 1.2*Vstall | 300 |
| H | Arrival Terminal Procedures | 1.2*Vstall | 300 |
| I | Transition + Descend | 1.2*Vstall to 0 | 40 |
| J | Hover Descend | 0 | 0 |
| K | Ground Taxi | 3 | 0 |

Source: Adapted from [39].

On-demand air taxi operations moving people between fixed or ad hoc locations are missions in which eVTOL vehicles fly from one point to another in an urban environment. This is a broad application of UAM resources once it may vary from short-length trips (e.g., within the city center) to trips between different cities.

Air cargo operations moving goods between warehouses and stores is a very similar operation to the aforementioned air taxi, in which the primary goal is to fly from one point to another. This type of mission is expected to solve problems faced by companies regarding logistics and product delivery.

Regularly scheduled "air metro" operations transporting passengers between a set of fixed locations are related to air taxi operations, but involves a more significant number of people. This is one of the possible applications of the Uber Elevate project [9].

Emergency medical evacuations, rescue operations, and humanitarian missions represent prioritized operations in which the Air Traffic Control (ATC) system is responsible for managing the airspace and for making it resilient.

Weather monitoring is another impacted activity once eVTOL vehicles can fly in urban environments and to collect more precise data regarding weather events (e.g., cloud formations). This could enable more precise real-time estimations and flow management adaptations. This mission can also be conducted as a part of the other mission (e.g., the delivery of people around the city can be conducted in parallel to weather information collection).

Ground traffic assessment can be useful for optimizing ground transportation systems in large-scale. In order to accomplish this, eVTOL vehicles can be employed to observe, collect the data, merge the data of different spots around the city, and provide people with real-time and accurate information.

Finally, the missions considered in this research involve delivering people from different points of the urban environment, which is represented by air taxi missions.

### 3.2.3 Hazard Scope

The operation of a new transportation system brings many advantages to society, but it comes with potential hazards that may lead the system to unsafe states and must be treated carefully. The authors in [24] list different examples of potential hazards for different UAM domains:

- · Vehicle, equipage, systems: Loss of electrical power and failure of position monitoring;
- · Vehicle servicing and maintenance: Unavailability of aircraft part for replacement;
- · Communications (voice datalink, Command-and-Control (C2) link): Failures in Command and Control (C2) Link[4] and reduced transmission Quality of Service (QoS) in critical situations;
- · Aerial operations, flight procedures, flight management: vehicles operating in altitudes that are not included in the pre-defined altitude interval;
- · Routing, airspace, air traffic management: UAM route conflicts with other types of air traffic (e.g., conflict with commercial aircraft in airport areas);
- · External environment (weather, obstacles, aerial traffic, birds): Adverse weather conditions;

---

[4]The Command and Control (C2) Link is a datalink that can be used to piloted aircraft remotely [41] [42].

- Pilots (onboard and remote): Lack of appropriate pilot training (from safety perspectives) and loss of situational awareness;
- Ground-based operations and infrastructure: Lack of vertiport (or skyport) availability;
- Cybersecurity: Problems regarding authentication and security aspects of C2 link.

## 3.3 Urban Air Mobility (UAM) Challenges

Furthermore, many technical challenges must be faced to achieve mature UAM operations. The authors in [24], [27] and [9] highlights a set of challenges for the next years:

- Congestion Management: this is a problem faced in dense transportation systems. However, methods to manage the scheduling and routing of eVTOL vehicles could enable safer and more efficient operations. Research efforts could include defining the routes (from strategical and tactical perspectives) and developing the schedules for departure and arrival at vertiports (or skyports);
- Scheduling: scheduling in the network is a challenge that needs to consider the characteristics of individual skyports (e.g., sequencing, scheduling, and spacing of UAM arrivals and departures). Optimized schedules can provide better safety, efficiency, capacity, and community acceptance;
- Separation: The separation standard is an area of research that might be driven by a pre-defined Target Level of Safety (TLS). However, different aspects that may be used for estimating the minimum separation between aircraft can be highlighted, such as the level of autonomy and the role of humans in the process.
- Interoperability: This challenge relies on the interaction between UAM components (e.g., eVTOL vehicles and ATC) and other users of the airspace (e.g., drones and general aviation). This is essential for the simultaneous operation of different technologies.
- Certification process: certification of eVTOL vehicles is essential for ensuring the capabilities of the aircraft to operate safely. Thisis a challenge forregulatorsto define the aspects and requirementsfor a valid certification process for this new technology;
- Airspace System Design & Implementation: This challenge is related to the construction of the UAM environment and could be inspired in methods applied into the general airspace. This includes operational rules, roles, procedures, and UAM skyports design.
- Community Integration: One important challenge is social acceptance. This is a crucial obstacle to be carefully faced once it is determinant to the success of the UAM initiative;
- Air Traffic Management (ATM): Many aspects of Air Traffic Management (ATM) are challenging in the UAM context and demand new technologies. For instance, Uber expects the ATC system to operate as a server that can deconflict global traffics based on the requests of the eVTOL vehicles;
- Air Traffic & Fleet Operations Management: This is also related to ATC. Organizations need to gather efforts from different areas in order to build safe, efficient, and scalable Air Traffic Management (ATM) operations considering new aspects (e.g., integration of self-piloted aircraft).
- Trajectory-Based Operations: The task of planning and managing trajectories to be precisely followed is challenging once the airspace is a complex environment that may change its state throughout the aircraft operation (e.g., emergencies and bad weather conditions). Furthermore, there is the challenge of ensuring the eVTOL vehicles do not follow conflicting trajectories once the entire operation is expected to included several vehicles.

Moreover, MITRE [29] presented a list of challenges faced in the UAM project that is also being faced in small Unmanned Aircraft System (UAS) operations mainly due to the development of the NASA's UAS Traffic Management (UTM) project [11]. The following paragraphs present some challenges faced by UAM and how the small UAS community have been facing them based on the report published by MITRE [29].

The airworthiness of eVTOL vehicles is the first point highlighted. The definition and classification of the technologies that make this aircraft to present unique flight profiles (e.g., Distributed Electric Propulsion - DEP [40]) do not fit within current classifications. For UAS, the industry and research communities have developed several technologies for enabling electric propulsion, for example. These technologies may serve as an inspiration for building new systems to eVTOL vehicles, i.e., to considerably larger aircraft configurations.

The certification process of autonomous flights is also complicated for these new systems. However, it is also a challenge for small UAS. Some applications of these small aircraft are conducted fully autonomously (e.g., package delivery) and some companies related to this market are dealing with the problem of building a certification system. Furthermore, this problem is also faced in the context of self-driving cars.

A large number of piloted and self-piloted eVTOL aircraft operating into the airspace will demand a high level of automation in Air Traffic Control (ATC). In this sense, UTM-like services could be a viable solution, once they are focused in low-altitude environments (below 400ft) and could be extended to higher altitudes (above 400ft).

In terms of detecting and avoid, procedures and standards need to be defined in order to highlight separation performance requirement. It is worth noting that eVTOL vehicles might be expected to detect obstacles (e.g., other flights) even if it is human-piloted. As detect-and-avoid standards are under development for small UAS using different technologies (e.g., radar and lidar), extensions for UAM aircraft will be needed.

In terms of operating rules, Instrument Flight Rules (IFR) are not intended to be employed into the UAM environment once it is expected to include too many aircraft, which harden the maintenance of current separation requirements. Thereupon, new technologies are needed in order to make self-piloted eVTOL aircraft able to operate considering IFR or new modified rules may be created. Furthermore, most of small UAS flights are conducted considering Visual Flight Rules (VFR)-like operations.

### 3.4 Trajectory-Based Operations (TBO) in UAM

Enhancing the airspace operations in general (considering both the National Airspace System - NAS - and UAM) is a challenging task once these operations take place into complex environments. Conversely, some strategies have been proposed for leveraging improvements in these operations. One approach proposed and that has been considered a viable solution for many problems currently faced is the Trajectory-Based Operations (TBO) concept. Thus, this Section presents the aspects related to TBO. Firstly, the definition of this concept is presented. Then, the main differences of the current airspace operation and TBO are highlighted. After that, the challenges regarding the implementation of TBO are presented. Finally, the challenges and possibilities of conducting TBO in UAM environments are shown.

### 3.4.1 Trajectory-Based Operations (TBO) Definition

Trajectory Based Operations (TBO) is the "exchange, maintenance and use of consistent aircraft trajectory and flight information for collaborative decision-making on the flight" [43]. TBO is a new operations concept that relies on increased predictability of aircraft 4D trajectories [44], enabling robust performance-based 4D trajectory management by sharing trajectory information [45]. Furthermore, TBO is an ATM method for strategically (rather than tactically) planning, managing, and optimizing flights that is supported by (i) Time-Based Management (TBM), which highlights the importance of deadlines for each activity/phase (e.g., departure and arrival); and (ii) Performance-Based Navigation (PBN), which describes the capabilities of the aircraft of navigating according to performance standards [46].

Once airspace congestion is impacting the airspace capacity and efficiency, new technologies are needed to plan, manage, and optimize flight trajectories strategically. In this sense, 4D navigation provides precise, cost-effective, and repeatable flight trajectories for enhancing the airspace operation [47].

"The scope of Air Traffic Management (ATM) in a TBO environment is to manage flights' trajectories and their interactions to achieve the optimum system outcome with minimal deviation from the user-requested flight trajectory" [48]. This requires a high automation level for all stakeholders in their decision-making processes [48]. In this concept, each aircraft is required to follow a (pre-)negotiated conflict-free trajectory. These trajectories, accurately planned in 4 dimensions (x, y, z, and t - time), reduce significantly the need of Air Traffic Controller (ATCo)'s intervention during the tactical phase (i.e., during the aircraft operation) [49] [50].

Moreover, TBO relies on five main features [44] [51] [52] [53]:

· Trajectory Management: Refers to the planning and execution of trajectories into the airspace operation. The task of planning and managing trajectories5 to be precisely followed is challenging once the airspace is a complex environment that may change its state throughout the aircraft operation (e.g., emergencies and bad weather conditions). Thus, the trajectory is classified into four types: Business Development Trajectory (BDT), which represents the trajectory that has been precisely planned; Shared Business Trajectory (SBT), which represents the trajectory that has been planned and sent to the ATM system and becomes available to other stakeholders; Reference Business Trajectory (RBT), which represents the negotiated trajectory that has been agreed by different stakeholders;

---

5which can start just before the flight starts or even months ahead of the launch date

· Collaborative Planing: Refers to the Collaborative Decision-Making (CDM) processes for satisfying, at a certain level, all stakeholders. This is supported by wide and structured data sharing and is useful for ensuring all stakeholders have acceptable solutions for their challenges;

· Integration of Airport Operations: Refers to the complete integration of airports into the ATM system. This enables, for example, global trajectories re-planning in case of problems in landing/takeoff procedures in specific airports;

· New Separation Modes: Refers to alternative ways of defining horizontal and vertical separation standards. These modes can be classified into three categories: Conventional aircraft separation, which is similar to the modes currently adopted into the airspace; Novel ground-based aircraft separation techniques, in which the aircraft are guaranteed to follow 4D trajectories in accordance with Required Navigation Performance (RNP) once the Precision Trajectory Clearance (PTC) is used; and Novel air-based aircraft separation techniques, in which the aircraft solve specific conflicts autonomously and collaboratively;

· SWIM: Stands for System Wide Information Management (SWIM) and refers to the data exchange, including all stakeholders of the process (e.g., ATC, ATCo, and airlines). The current plans expect the future information system to meet the following requirements: Standardized formats; Data availability to different users; high quality and high relevance data.

TBO also considers precise predictions (position and time) a flight will be throughout its operation (i.e., it is focused on strategic planning). This enables the system to achieve higher throughput (using time-based management techniques), higher predictability (considering more accurate planning and scheduling processes), higher efficiency (more efficient flows), and higher operator flexibility (through increased user collaboration) [46].

In comparison to current ATM systems and approaches, TBO is expected to enhance planning and execution of efficient flights, reduce potential conflicts, and resolve network imbalances early. For this, the entire flight is considered: the planning process, the execution, and the post-flight activities [45].

### 3.4.2 Differences between the current airspace operation and TBO

In order to implement TBO into the airspace, an analysis of the differences of current and future system must be conducted. The authors in [44] presents an extensive discussion on these differences. The main Differences are:

· Airspace User Operations:
  – Instead of dealing with routes, future operations will focus on trajectories, which can be constantly updated during the flight, and agreed with all stakeholders;
  – Advanced navigation and data-link communication equipment will be required;
  – Self-separation will move some ATCo workload from the ATCo to the aircraft crew.

· Airport Operations:
  – Collaborative Decision-Making (CDM) will enhance flights and operations predictability;
  – Runways and taxiways will be used more efficiently;
  – Advanced control systems will reduce the ATCo workload and increase airport capacity in low visibility conditions;

· Airspace Organization:
  – Sectors are expected to be more flexible;
  – Dynamic allocation for special purposes (e.g., military operations) will be conducted;
  – Functional Airspace Blocks (FAB) will reduce the airspace fragmentation;

· Network Operations:
  – Strategic flow management are expected to be connected with tactical ATC operations by Trajectory Management;
  – Predictions of trajectory will be more accurate than it currently is;
  – The ATCo workload will be decreased once the deconfliction process is expected to be conducted strategically rather than tactically.

· Automated Systems:
  – Automation is expected to perform some ATCo tasks, which can increase the capacity;
  – Automation is expected to assist ATCo and pilots in the decision-making process;
  – Automation will be required in the aircraft avionics for self-separation purposes;

### 3.4.3 Challenges of the implementation of TBO in UAM environments

Although the relatively short distance and duration of UAM flight remove many of the obstacles faced in the TBO implementation, there still a set of challenges related to these operations [54].

One example of obstacle faced in UAM environments regarding the implementation of TBO is the separation standard. Current configurations of Instrument Flight Rules (IFR) are not compatible with the requirements of UAM, i.e., the separation adopted does not fit into the dense operations considered [26]. Conversely, one possible solution for reducing the separation standard of IFR UAM operation relies on the use of new Communication, Navigation, and Surveillance (CNS) technologies. These advancements can be used to enable the whole system to deal with precisely defined 4D flight trajectories [55]. This reduced separation approach can be followed "when special electronic, area navigation or other aids enable the aircraft to closely adhere to their current flight paths" [56]. One example of a similar and successful approach is the Performance-Based Navigation (PBN) Required Navigation Performance (RNP), which relies upon satellite-based navigation, ground-based and satellite-based augmentation systems (e.g., GBAS), and onboard performance monitoring and alerting to provide high operational integrity degrees [57].

Furthermore, planning, executing, and analyzing UAM flights is not a simple task [44]. As operations are expected to consider a large number of flights, trajectory management becomes complex. Indeed, the variability of the airspace in terms of aircraft numbers (i.e., different flights may be started and, then, be fully executed in a small period) highlights the challenges of dealing with trajectory management. In this sense, three requirements can be highlighted for the process of planning trajectories: (i) safety (i.e., the trajectory must be conflict-free); (ii) efficiency (i.e., the trajectory must conduct the aircraft to execute its flight as quickly as possible); (iii) response-time (i.e., the trajectories are expected to be planned in a short period of time, even considering a complex airspace). Finally, updating the trajectory throughout the flight is another challenge in UAM operations.

Moreover, Collaborative Decision-Making (CDM) is another challenge in this context [44]. In order to satisfy (at a certain level) all stakeholder of UAM operation (e.g., authorities, airlines, and users), relevant information needs to be shared. Indeed, there are challenges regarding standardization of information sharing and approaches for involving multiple stakeholders in key decisions. For example, the communication between NAS operators and UAM operators can be considered a challenge once they might need to work collaboratively for finding satisfactory solutions to the problems they might face.

Finally, another example of challenge faced in the TBO implementation in UAM environments refers to the integration of skyports into the data management system [44]. The process of takeoff and lading can vary considerably depending on the skyport characteristics (e.g., location and altitude). Furthermore, the ground-taxi can also be considerably different depending on other aspects (e.g., skyport capacity). Thereupon, sharing information regarding real-time skyport operations is essential for enabling more accurate trajectories and scheduling planning. However, there are aspects of these operations that harden the precise estimation of the phases conducted in this region, such as data standardization, aircraft characteristics (i.e., different aircraft models may perform takeoff and landing differently), and weather forecasting.

## 4 Trajectory-Based UAM Operations Simulator (TUS)

In this Chapter, the Trajectory-Based UAM Operations Simulator (TUS) is depicted. This simulation tool is coined for evaluating trajectory-based UAM operations and is extensively dicussed in this Chapter. In fact, the simulation of such environment is essential for enabling appropriate planning methods and operational schemes. In this sense, it is important to define the goals, scope, and scenarios the simulation tool is intended to deal with. Firstly, the definition of this proposal is presented. Secondly, the principles and the assumptions considered are presented. Then, the simulation process is discussed. Finally, the implementation of TUS is described.

### 4.1 Definition

The main goal of the Trajectory-Based UAM Operations Simulator (TUS) is to simulate the Trajectory-Based UAM operations in urban environments considering the presence of both manned and unmanned eVTOL vehicles. For this, a Discrete Event Simulation (DES) approach is adopted, which considers an input (i.e., the eVTOL vehicles, their origin and destination, and their respective trajectories) and produces an output (which describes if the trajectories are safe and the elapsed operation time). The main contribution of this simulation tool is to provide a simulated environment for testing and measuring the effectiveness (e.g., flight duration) of trajectories planned for eVTOL vehicles.

This simulation tool is based on the Final Sector Simulation Tool (FSST) implementation, but it also is inspired in different simulation tools (e.g., BlueSky [30] and TAAM [58]). Although this tool is intended to be used as a server-side application (i.e., it is intended to be used by urban airspace managers for enhancing the system performance), future implementations might also consider the usage of such proposal as a client-side application. This could enable a more collaborative relationship between end users and urban airspace manager and regulators.

This first step towards the simulation of Trajectory-Based UAM operations is vital from a long-term perspective. Indeed, the contributions of this research can leverage the production of new tools for enhancing UAM operation, both extensions of we aim to do herein and different research aspects (e.g., capabilities of electric batteries).

Finally, this simulation tool considers a set of assumptions that are explained in the following sections. This highlights that this proposal is tailored for dealing with the current research challenges faced with UAM context. Thus, future proposals could include a different set of assumptions into the UAM operation for dealing with different problems. The advancements in UAM simulation that may be developed in the following years can leverage the research community to push the boundaries faced nowadays and to enable the creation and development of precise procedures that consider all aspects of this operation.

## 4.2 Goal and Principles

The Trajectory-Based UAM Operations Simulator (TUS) aims at simulating Trajectory-Based UAM operations considering multiple Electric Vertical and Take-off Landing (eVTOL) vehicles in urban environments. For this, a Discrete Event Simulation (DES) environment is developed for testing and measuring the effectiveness of different aspects regarding trajectory management. Thereupon, there are a set of principles considered that provides a better understanding of the scope, assumptions, and capabilities of this simulation tool. These principles are:

1. Environment for building UAM trajectories: The simulation ecosystem is intended to provide a well-suited environment for planning eVTOL vehicles trajectories from taking off to landing. However, this tool is not intended to simulate all aspects of the airspace. For instance, the power consumption of each trajectory and intersection between UAM and other regions of the National Air System (NAS) are not included in our scope;

2. Feasibility of trajectories: The trajectory planned must respect the Electric Vertical and Take-Off (eVTOL) vehicles specs, i.e., the trajectories planned must be feasibly accomplished by these vehicles accordingly to performance specs (e.g., speed range and climb/descent rates);

3. Horizontal and Vertical Separations: In this research, horizontal and vertical separations are intended to be respected regardless of the altitude. Thereupon, two eVTOL vehicles are allowed to fly into the horizontal position if a proper vertical separation ($z$) is applied. These principles ensure there is a considerable sparsity among vehicles according to the specialists consulted. As illustrated in Figure 2, the separation between two eVTOL vehicles can be represented as a cylinder that considers the whole cruise altitude interval;

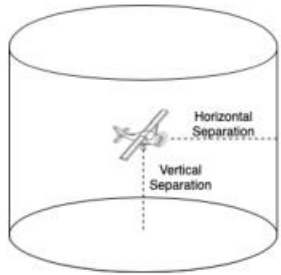

Figure 2: Horizontal and Vertical Separations of eVTOL vehicles.

Source: The author.

4. Trajectory-Based Operations: TUS considers that the trajectories planned for eVTOL vehicles are precisely followed. In fact, the trajectories provided as an input for the eVTOL vehicles drive their operation and conflicts are not intended to be solved tactically, i.e., conflict-free trajectories are intended to be built strategically. This highlights that TUS is aimed at measuring if a set of trajectories provided are safe (i.e., if it respects the minimum separation standards) and efficient, and not to measure the capabilities of the eVTOL vehicles (or ATC) to avoid/solve conflicts tactically.

5. Trajectory Definition: Once the longitudinal and vertical separation are respected regardless of the altitude of the vehicles, this research considers trajectory as a tuple $t$ composed of longitudinal coordinates ($x$, $y$), altitude indication ($z$), and speed at which the vehicle must intercept the position ($s$). From this, the time os intersection can also be estimated. A trajectory with $n$ positions is illustrated in Equation 1.

$$t = [(x_1, y_1, z_1, s_1), (x_2, y_2, z_2, s_2),..., (x_n, y_n, z_n, s_n)] \tag{1}$$

Moreover, note that the trajectory can be formatted according to TBO requirements when the simulation is finished, i.e., interception time can be estimated rather than speed of interception.

6. Instrument Flight Rules (IFR): The trajectories built in this research assume the eVTOL vehicles can operate accordingly to Instruments Flight Rules (IFR), i.e., the aircraft must follow the trajectory defined as precisely as possible. Achieving the airspace density required by mature UAM operations is infeasible if current IFR separation standards are considered [26]. Conversely, additional opportunities for reducing IFR separation standards have been identified[6] [54]. Indeed, a set of potential approaches for reducing the minimum separation of IFR operations focus on the use of new Communication, Navigation, and Surveillance (CNS) technologies. These approaches rely on precisely defined 4D flight trajectories and the assignment of aircraft to reduced-separation trajectories [55]. One example of such operation is the Performance-Based Navigation (PBN) Required Navigation Performance (RNP)[7] [57]. Thereupon, it is reasonable to consider that IFR aircraft can fly with reduced separations in future low-altitude operations (especially in future Trajectory Based Operations (TBO) [59][8]. Finally, a future direction of this research is the measurement of suitable separation standards considering the environment faced[9];

7. UAM Maturity Level (UML): Although the simulation tool presented herein can be scaled to consider multiple regions simultaneously in future works, our focus is to simulate an area that reasonably represents the urban area of one city. Thereupon, our proposal is designed to propose solutions for different maturity levels (illustrated in Table 1): UML-1 (for very initial operations with a low number of vehicles), UML-2 (which considers a small UAM network), UML-3 (in which the operations are conducted into urban core), and UML-4 (in which many vehicles are able to operate simultaneously). However, UMLs 5 and 6 consider the simultaneous operation of hundreds and thousands of vehicles, which is infeasible for only one city. However, one possible future direction relies on the extension of this simulation tool for dealing with multiples regions simultaneously;

## 4.3 Assumptions

Simulating UAM operations is a challenge once there are several aspects to be considered (e.g., human-machine interaction, machine-machine interaction, skyports, and the intersection between UAM and the National Airspace System - NAS). To build a simulator for analyzing specific aspects, a set of assumptions are adopted. In this Section, we present all assumptions made for building the Trajectory-Based UAM Operations Simulator (TUS). Firstly, the assumptions regarding the airspace constraints are highlighted. Then, the assumptions on the mission scope considered in this research are depicted. Finally, the assumptions on the Electric Vertical Take-Off and Landing (eVTOL) vehicles are presented.

### 4.3.1 Airspace Constraints

The assumptions of airspace constraints focus on the characteristics of the operations conducted. Three main aspects are highlighted: Flight Rules, Boundaries, and Capacity.

Flight Rules The trajectories built in this research assume the eVTOL vehicles can operate accordingly to Instruments Flight Rules (IFR). As additional opportunities for reducing IFR separation standards have been identified [54] focusing on the use of new Communication, Navigation, and Surveillance (CNS) technologies, precisely defined 4D flight trajectories stands as the foundation of this assumption [55]. One example of reduced-separation trajectories that can be used as an inspiration for IFR procedures into UAM operations is the Performance-Based Navigation (PBN) Required Navigation Performance (RNP) [57]. Hence, it is reasonable to consider that IFR aircraft can fly with reduced separations in future low-altitude operations.

Horizontal and Vertical Boundaries The UAM project is designed for solving transportation problems within urban environments. Although there is a potential large-scale usage of this new proposal (e.g., UAM operation across states), its early operations are expected to take place within cities. In this case, the area of operation assumed in this research

---

[6]The current challenges regarding separation reduction rely on specific factors, such as cost and implementation [54].

[7]PBN-RNP relies on satellite-based navigation, ground-based and satellite-based augmentation systems, and onboard performance monitoring and alerting. Thus, this is intended to provide high degrees of operational integrity [57].

[8]These operations rely on dynamically defined 4D flight trajectories with reduced containment boundaries [55].

[9]Note that the simulation tool may also be useful in measuring the impacts of different separation standards

refer to a reasonable size of the metropolitan area of a considerably large city. Then, we assume an interval of $0NM$ up to $30NM$ for both x- and y-axis. In other words, this covers an area of $900NM2$, which represents a considerably sizeable urban area. Conversely, future implementations of this simulator may be intended to address the issue of dealing with extensive UAM operations (e.g., operations across different countries), but these operations are out of the scope of this research. Concerning the vertical boundaries, UAM operations are expected to be conducted in an altitude similar to 1000ft AGL [39]. Thereupon, cruise operations are conducted in four different altitudes, which are faced herein as flight levels: 1000ft, 1200ft, 1400ft, and 1600ft10;

Capacity The amount of traffics the airspace is able to maintain is a research topic with different approaches [60][61][62]. The main goal in capacity estimation is to maintain the airspace efficient (i.e., delivering as many aircraft as possible) without compromising the safety levels [63]. Once the focus herein is to simulate UAM operations, we assume the flow management (i.e., the amount of aircraft that will be taking-off in a given time window) is out of our scope. In other words, the proposed approach aims to measure if trajectories provided beforehand can safely conduct a specific fleet, from origin to destination, regardless of the amount of traffic to be delivered. Indeed, if the number of traffics is considerably large, the proposed platform is likely to show that the trajectories proposed are not safe due to the physical limitation (e.g., ten eVTOL vehicles in a minimal area are likely not to respect the separation standards) and the minimum separation requirements. This shows that the scope of the Trajectory-Based UAM Operation Simulator (TUS) relies on measure how well trajectories fit the requirements (i.e., safety and efficiency). Furthermore, a possible direction of this research is the automatic Air Traffic Flow Management (ATFM) based on the outcomes of TUS for enhancing, for example, scheduling and load balancing;

### 4.3.2 Mission Scope

The assumptions of the mission scope present characteristics of the missions conducted. Three main aspects are highlighted: Transportation, Skyports, and Flight Phases.

People Transportation The main goal of the UAM initiative is to enhance the transportation system into urban environments. Indeed, there are different mission scopes included in this initiative, such as on-demand air taxi operations, air cargo operations, regularly scheduled air metro operations, emergency medical evacuations, weather monitoring, and ground traffic assessment [24]. These missions are considerably different from the requirements perspective. Conversely, the mission scope of this research focuses on on-demand air taxi operations, in which eVTOL vehicles fly from one specific position (i.e., the origin skyport) to another specific position (i.e., the destination skyport). Although there is a recommendation on the maximum distance between these two points [9], we consider that the vehicles can perform this operation within the urban environments (i.e., $30NM \times 30NM$) since a large aircraft autonomy is assumed;

Skyports Skyports can be faced as the airports of eVTOL vehicles in UAM environments. A consistent format is to operate on the top of constructions [9], although it can also be considered on the ground. Figure 3 illustrates the skyport assumed in this research. The vehicles land/take-off at/from specific constructions that can offer an appropriate infrastructure. Furthermore, the cruise area in this research is assumed to be around 1000ft Above Ground Level (AGL) [39], whereas the skyport is assumed to be 100ft AGL. Finally, the skyport capacity is assumed to be considerably large11;

Flight Segments UAM operations consider different flight segments. Basically, on-demand air taxi operations present 11 segments [39] that can be divided into three categories:

1. Departure Segments (ground taxi, Hover Climb, Transition and climb, Departure terminal procedures, acceleration and climb );

2. Operation Segments (cruise);

3. Arrival Segments (deceleration and descend, arrival terminal procedures, transition and descend, hover descend, ground taxi).

Thereupon and given the goals of the Trajectory-Based UAM Operations Simulator (TUS), the simulations are focused on (i) simplified take-off procedures, (ii) cruise trajectories, and (iii) simplified landing procedures. This highlights that a simple procedure is adopted in take-off/landing, although more elaborate procedures can be defined in future

---

101000ft and 1400ft are considered odd Flight Levels, whereas 1200ft and 1600ft are considered even Flight Levels.

11In this research, skyports are assumed to support the presence of multiple eVTOL vehicles simultaneously, i.e., the capacity of these areas are reasonably compatible with the number of vehicles flying around the urban area.

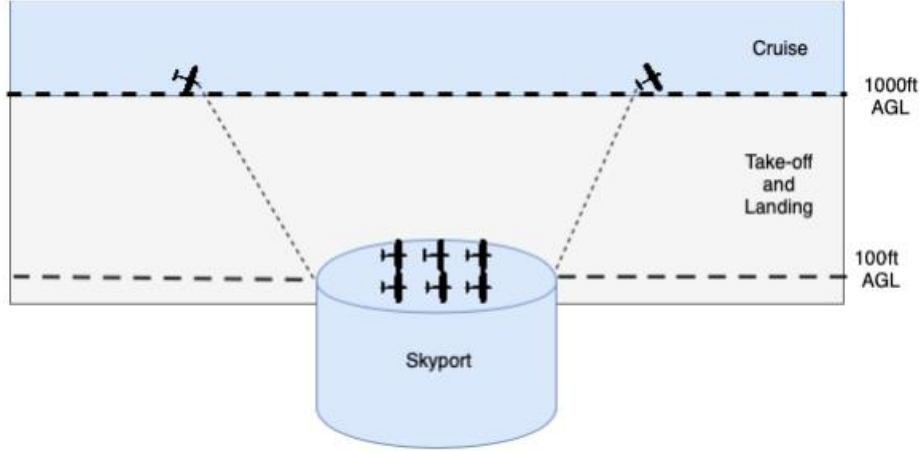

Figure 3: Skyport in UAM operations including take-off an landing areas and cruise area.

Source: The author.

works. Finally, we assume the challenge of conducting the vehicles for safe and efficient landing and specific phases of take-off/landing procedures (e.g., ground taxi) are not included in our scope.

### 4.3.3 Electric Vertical Take-Off and Landing (eVTOL) Vehicle

The eVTOL vehicle assumptions include different aspects of the operation of this aircraft. Although there are other manufactures-dependent aspects (e.g., aircraft design), the focus here is to present the main characteristics in specific performance aspects. Hence, six main aspects are highlighted: speed, rate of turn, piloting classification, minimum separation, cruise condition, and further specs.

Furthermore, we consider two types of autonomous eVTOL vehicles in this research: Remotely Piloted Aircraft System (RPAS) and Self-Piloting eVTOL vehicles (also referred as UAS in the context of this research). The RPAS stands for a system with an operator, i.e., the aircraft is not entirely autonomous12. In this system, the Command and Control (C2) link connects the Remote Pilot Station (RPS) and the RPAS for controlling the flight [64]. Furthermore, Autonomous Aircraft (AA) are piloted by software without human operators intervention. Thereupon, the operation of piloted and self-piloted eVTOL vehicles are comparable, i.e., we assume RPAS and AA are capable of following recommended trajectories.

Speed Several manufacturers are developing different eVTOL models [35] [36] [37] [38]. Although each model has its performance specs and configurations, a reasonable en-route eVTOL vehicle airspeed varies from 130kts to 170kts at 1000ft Above Ground Level (AGL) [9] [40]. In this research, we assume a speed interval of 130kts-170kts for en-route eVTOL vehicles operations [9]. Instead of simulating precisely the kinematics of the eVTOL vehicles, our goal in the simulation process is to provide an operational profile that is compatible with the eVTOL requirements [39]. Considering this speed interval enables (i) vehicles to follow the trajectories with a lower uncertainty once this speed relies upon between the expected eVTOL vehicle performance range, (ii) the system to be more predictable once arrival estimations become simpler, and (iii) easier adaptability of vehicles regarding the speed. Finally, acceleration and deceleration are other two important parameters in this operation. Considering the aircraft specs, the type of operation conducted, the speed range, the requirements presented by Uber [39], and the opinion of specialists in Air Traffic Management (ATM), reasonable acceleration and deceleration values are, respectively, 1kts/s and 2kts/s. Conversely, this value can be easily adjusted in future works, and our goal herein is to simulate initial and standard scenarios.

Rate of Turn The dynamics of the heading variation changes considerably, depending on the model of the vehicle. In the National Airspace System (NAS), the standard rate of turn is 3 degrees per second [65]. Indeed, the aerodynamics of these aircraft are different in comparison to the models intended to be used in UAM operations. For helicopters, this rate of turn tends to be considerably higher once the operations are also different. Thereupon, a reasonable maximum rate of turn relies on around 7.2 degrees per second once the airspeed considered is close to 150kts [66]. Hence, we assume the eVTOL vehicles are capable of changing their respective heading up to 7.2 degrees per second in all the speed spectrum (130kts-170kts) once this is a considerably small speed variation. Furthermore, it can be adjusted in future works considering the detailed performance description of specific eVTOL vehicles models.

---

12The International Civil Aviation Organization (ICAO) has been working to integrate the RPAS in terms of an international regulatory framework [64]

Climbing and Descending rates

Another assumption regarding the eVTOL vehicle operation refers to the climbing and descending rate. In [39], the authors highlight that in all flight phases, the maximum climbing rate and the descending rate is 500ft/min. Indeed, this is a viable starting point once it represents the requirements for operational vehicles. Thereupon, we consider a maximum climbing and descending rate of 500ft/min. Finally, we consider a simplified approach in which the aircraft can turn the bank to an angle that achieves 500ft/min in one second. This highlights the aircraft inclination may be treated differently in future works.

Piloting Classification

Although early UAM operations are expected to consider only piloted vehicles[13], Uber is expecting that "vehicles will have the avionics and sensors needed for autonomous flight" [39]. Furthermore, we consider two types of autonomous eVTOL vehicles in this research: Remotely Piloted Aircraft System (RPAS) and Self-Piloted eVTOL vehicles. The RPAS stands for a system with a human operator, i.e., the aircraft is not fully autonomous[14]. In this system, the Command and Control (C2) link connects the Remote Pilot Station (RPS) and the RPAS for controlling the flight [64]. Furthermore, Self-Piloted eVTOL vehicles are piloted by software without human operators intervention. Thereupon, the operation of piloted and autonomous eVTOL vehicles is comparable, i.e., we assume RPAS and self-piloted eVTOLs are capable of following recommended trajectories, even considering the present lack of social acceptance.

Minimum Separation As UAM operations are expected to be conducted following reduced longitudinal separations, the development of new technologies for enabling such operations is demanded. When procedures follow current IFR primitives, one reasonable minimum separation is 3NM [54]. Conversely, this separation is not feasible for dense urban operations. In [67], the authors performed different simulations with reduced separations in UAM environment. Thereupon, our proposal assumes 0.25NM as the minimum horizontal separation for considering the specialists' opinion and the range of values adopted in [67]. Moreover, although autonomous vehicles may bring benefits to the airspace operation [68] [69] [70] [71] [72] [73], there are many challenges regarding their operation. One important challenge in this context is the social acceptance [74]. Thereupon, the minimum horizontal separations for RPAS and self-piloted vehicles are assumed to be higher than the separation of piloted vehicles. The separation assumed herein for both types is 0.5NM, which is considerably larger than the aforementioned 0.25NM and provide these autonomous vehicles with more time for decision-making in critical situations (e.g., emergencies). Regarding vertical separation, piloted and autonomous vehicle requires a minimum separation of 200ft. This separation is assumed based on the aircraft characteristics and the considerably short airspace slice of UAM operations. Finally, future works may also employ different vertical and horizontal separations based on future regulations and different vehicle specs.

Cruise Condition In UAM operations, cruise segments are expected to be conducted in an altitude range close to 1000ft Above Ground Level (AGL) [39]. However, this requirement does not exclude vehicles crossing considering different altitudes, i.e., two vehicles may be allowed to be at the same latitude and longitude, at the same time, if the altitude difference between them is considerably large. In this research we consider four flight levels: 1000ft, 1200ft, 1400ft, and 1600ft. Similarly to the approach adopted into the National Airspace System (NAS) [75], this research considers different flight levels for different headings. Figure 4 highlights the cruising levels variation regarding the heading applied. If the aircraft heading is less than 180 degrees, it is allowed to operate with altitudes of 1000ft and 1400ft. Otherwise, the operation can be conducted with altitudes of 1200ft and 1600ft.

Furthermore, successful operations in the first years of UAM operations are vital for long-term deployment of urban air transportation. Thereupon, this research considers that the longitudinal distance requirement by each aircraft (i.e., 0.25NM for piloted and 0.5NM otherwise) must be applied when the minimum vertical separation is not followed for ensuring there is a space between the aircraft.

Further Specs

Although the eVTOL vehicles are expected to solve several issues of the existing transportation systems (e.g., congestion), there are some limitations that have to be handled by UAM operators. One of the challenges regarding eVTOL vehicle operation is autonomy. As these aircraft are battery-powered, there has been an effort of several companies in developing advanced energy storage system for enhancing the UAM operation. The distance an eVTOL vehicle is

---

[13]This is for assuring autonomous systems are capable of operating safely and efficiently based on the data that will be collected.

[14]The International Civil Aviation Organization (ICAO) has been working to integrate the RPAS in terms of an international regulatory framework [64]

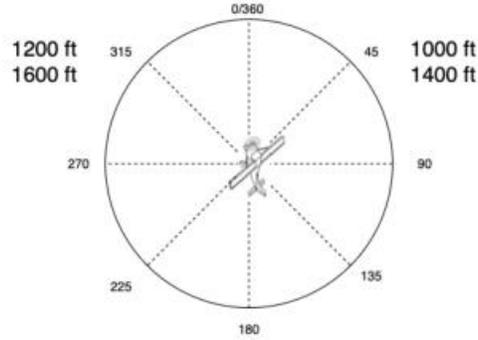

Figure 4: Table of Cruising Levels considered in this research.

Source: Adapted from [75].

capable of flying depends considerably on its design. For instance, the authors in [39] state that the vehicles should be able to fly 60 miles (or 52.14 Nautical Miles). In this research, we assume the eVTOL vehicles are capable of flying from any position to any other position in the urban environment. This enables us to plan trajectories for future vehicles with extended autonomy without compromising the trajectory planning service for initial aircraft. Hence, note that the urban environment considered in this research is represented by a square of size $30NM \times 30NM$. Furthermore, the eVTOL vehicle is assumed not to exceed 45' in its max dimension, and the max takeoff weight and not to exceed 7,000 lbs15.

### 4.3.4 Summary

In this Section, we presented the assumptions considered in this research. Table 3 summaries all considerations discussed.

Firstly, the assumptions on the airspace constraints were presented (flight rules, boundaries, and capacity). Then, the mission scope was depicted (people transportation, skyports, and flight segments). Finally, the aspects of eVTOL vehicles were shown (speed, rate of turn, piloting classification, minimum separations, cruise conditions, and further specs). These assumptions support the definition of the simulation tool (presented in the following Chapter) and the proposed trajectory planning platform.

## 4.4 Simulation Process

The Trajectory-Based UAM Operations Simulator (TUS) implementation structure is presented in Figure 5. The UAM_manager object is responsible for conducting the simulation process as it verifies if the airspace is safe and conduct the aircraft throughout their pre-defined trajectories. This object contains a set of eVTOL vehicles that represent the current airspace state. The eVTOL object represents the eVTOL vehicle, which follows the assumptions of this research and presents a pre-defined trajectory (which could be safe and efficient or not). Finally, the VideoMaker object is connected to the UAM object for building videos that enable a visual understanding of the trajectories followed. Hence, TUS acts as a platform for enhancing the trajectory planning process once it measures the effectiveness of each solution.

---

15This definition highlights that small aircraft (e.g., package delivery drones) are not considered in our scope

Table 3: Recap on the assumptions adopted in this research.

| Assumption | Topic | Consideration |
|---|---|---|
| Airspace Constraints | Flight Rules | IFR-like |
| | Boundaries | 30NM x 30NM |
| | Capacity | Limited by vehicles' separation |
| | People Transportation | On-demand air taxi operations |
| Mission Scope | Skyports | considerably large capacity |
| | Flight Segments | Take-off, Cruise, Landing |
| | Speed | 130kts-170kts |
| | Acceleration/Deceleration | 1kts/s |
| | Climbing/Descending rate | 500ft/min |
| | Rate of Turn | 7.2 degrees per second |
| | Piloting Classification | Piloted, RPAS and Self-Piloted |
| | Horizontal Separation | 0.25NM (Piloted) and 0.5 (Self-Piloted) |
| | Vertical Separation | 200ft |
| | Cruise Condition | 1000ft, 1200ft, 1400ft 1600ft |
| eVTOL Vehicle | Further Specs | Autonomy for flying in urban environments, <=45'in its max dimension, <=7000lbs |

Source: The author.

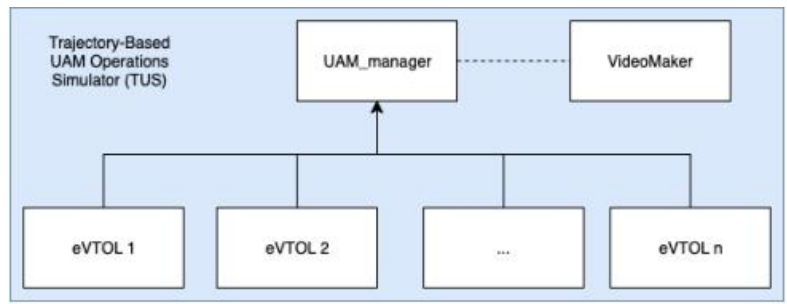

Figure 5: Implementation structure regardless of the simulation approach.

Source: The author.

Furthermore, Figure 6 shows a class diagram that provides a more in-depth explanation of how the TUS is implemented. In this diagram, the relationships of all classes are presented. Firstly, UAM_manager is responsible for modeling the current airspace scenario. The methods presented herein highlight features needed for the simulation process (e.g., check_conflict and filter_region_eVTOL). Furthermore, an UAM_manager instance is composed of different eVTOL vehicles, which represent the vehicles that have been assigned to a trajectory[16]. Then, the goal of the UAM_manager is to simulate the movements of these eVTOL vehicles in order to provide external applications with a broad understanding of the airspace state throughout the time.

The UAM_manager is, then, composed of eVTOL vehicles but also presents a relationship with VideoMaker. This static class provides the UAM_manager class with the capability of building videos from simulated scenarios for providing users with a visual understanding of the airspace operation. This feature is exciting as a human-machine interface once it is more intuitive to understand the trajectories. Conversely, it does not compose the system of building/simulating trajectories for UAM operations, i.e., it is just used for visualization and may not be required in production. Moreover, visualizing the vehicle movements may help managers and regulators to have a better situational awareness.

---

[16]This trajectory could be a direct flight or not.

Finally, the eVTOL class aims to model specific aspects of the eVTOL vehicle operation. All the attributes and methods presented herein refer to the vehicle movement and its capabilities of following new objective points. It also presents a common way of simulating one step in its operation (though the step method), similarly to the UAM_manager class.

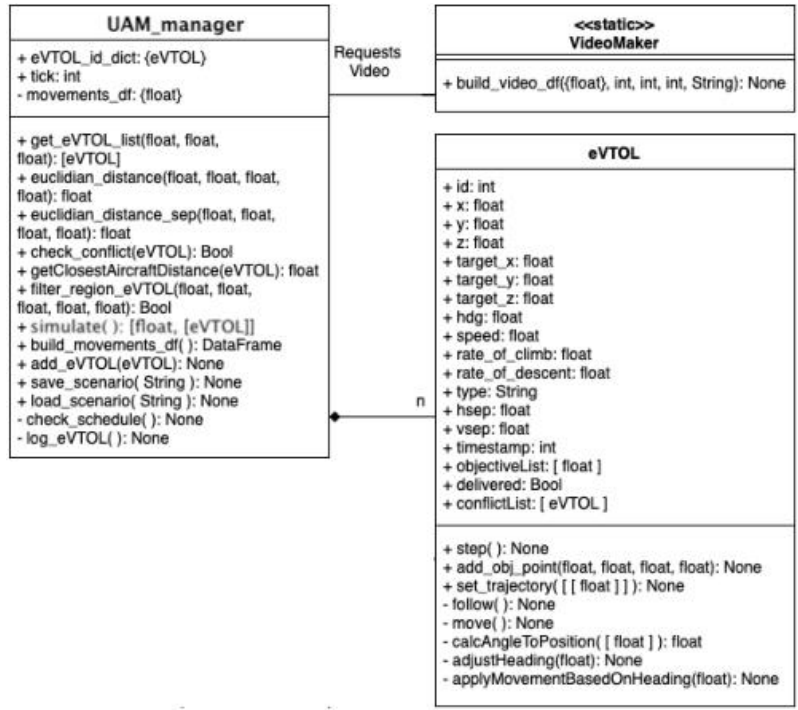

Figure 6: Class Diagram of the Trajectory-Based UAM Operations Simulator (TUS)

Source: The author.

.

## 4.5 Trajectory-Based UAM Operations Simulator (TUS) Implementation

In this section, we describe the responsibilities and characteristics of the TUS implementation. Firstly, the UAM_manager class is described. Then, the eVTOL class is depicted. Finally, the VideoMaker class is presented.

### 4.5.1 UAM_manager

The main goal of this class is to represent a given scenario considering an airspace state. In other words, this class represents the airspace and controls all the vehicles that have been assigned to a trajectory. Indeed, the trajectories of all vehicles must be respected in order to maintain the airspace safe. This highlights the importance of planning trajectories of new flights respecting the trajectories that have already been planned.

This class contains a dictionary that stores all the eVTOL vehicles using their ids as keys. Furthermore, the simulation time is stored in the *tick* variable. In each second of the simulation, the movements of all vehicles are stores into the movements_df variable. This variable represents a dictionary that logs all movements perceived in the simulation process. Finally, this dictionary is converted into a Data Frame afterward and, then, used by the VideoMaker object for building a trajectory visualization.

This class presents a considerable number of methods for performing specific tasks. Each method is explained as follows:

   · add_eVTOL: This method adds one eVTOL vehicle to the UAM object considering that its trajectory has been planned;

   · set_trajectory: This method sets the entire trajectory of the eVTOL vehicle in only one call;

· check_schedule: This is an internal method that checks if there is a scheduled eVTOL vehicle to take-off in a specific time;

· get_eVTOL_list_region: Given a specific position (x, y), this method returns all eVTOL vehicles present in a given radius (r);

· euclidian_distance: A simple implementation of the Euclidean distance equation;

· euclidian_distance_sep: An implementation of the Euclidean distance equation that includes the minimum separation of the eVTOL vehicles. The distance of two vehicles is given by the Euclidian distance subtracted by the minimum separation requirement of the most restrictive vehicles. For instance, if the Euclidian distance of two eVTOL vehicles is 5NM and the minimum separation of the vehicles are 0.25NM and 0.5NM, this method returns 4.5NM (5NM - max(0.25NM, 0.5NM));

· check_conflict: This method identifies if there is a conflict given an eVTOL as input. This computation involves the identification of the given and the surrounding vehicles and their respective minimum separation[17];

· getClosestAircraftDistance: This method returns the distance between a given eVTOL vehicle and its closest surrounding vehicle;

· filter_region_eVTOL: This method returns a list of eVTOL vehicles that close to a particular position (given a radius r);

· simulate: This method is responsible for conducting the simulation. One execution of this method represents a several seconds of operation in the simulation process: all eVTOL vehicles that have already been planned (i.e., the vehicles that are stored into the UAM object) move toward their goal. As 1 second is considered at a speed of 150kts, the vehicles fly 0.0417NM (as 1kt is equivalent to 0.000278NM per second, $150 \times 0.000278 = 0.0417$NM) in each second. The return of this method is composed of (i) the time spent to deliver all flights (this duration is considered infinite if the solution is not safe), and (ii) a list of conflicting eVTOL vehicle (if the entire operation is conflict-free, a empty list is returned);

· save_scenario: This method saves the whole UAM structure (i.e., all eVTOL vehicles and their attributes) into a JSON file. This file can be used for loading the same scenarios in other experiments. The JSON file is composed of a list of eVTOL vehicles that are described by the following attributes: "eVTOLid" (the id for identifying the eVTOL vehicle), "initial_position" (the first position of the aircraft in the simulation), "final_position" (the final position of the aircraft), "hdg" (initial heading of the aircraft), "eVTOL_type" (type of the eVTOL vehicle regarding the piloting process), "objectiveList" (a list of positions that the aircraft must achieve before going to its final position, i.e., the aircraft trajectory), and "timestamp" (the time in which the operation of the aircraft starts into the airspace).;

· load_scenario: This method uses the data stored into JSON files for building a pre-defined scenario. This feature is useful for enabling the communication of the simulation tool with external systems;

· log_eVTOL: This method adds the movement description of a given eVTOL vehicle into the movements_df attribute, i.e., it includes the trajectory of an external eVTOL (whose trajectory is being planned) to be displayed into the video;

· build_movements_df: This method adds the description of the current movements of all vehicles into the history of the movements. This is useful for building videos after when the simulation process is finished.

The simulation process is intended to be conducted using the *step* function, which represents one second of the aircraft operation. Finally, the strategy adopted to check if the simulation execution is finished is to check the distance of the vehicles and if there is at least one aircraft in operation or scheduled for operation. In other words, the simulation is considered to be finished (i) when all aircraft are delivered or (ii) when there is a conflict in operation.

### 4.5.2 eVTOL

In this simulation tool, the eVTOL vehicle state is represented by three metrics (illustrated in Figure 7): heading, speed, and position. The heading can be defined as "provision of navigational guidance to aircraft in the form of specific headings, based on the use of radar". It represents the angle the aircraft operates. Although the standard rate of turn of large aircraft operating into the National Airspace System (NAS) is 3 degrees per second [65], the aerodynamics of vehicles intended to be used in UAM operations are different. For helicopters, for instance, this rate of turn tends to be considerably higher than 3 degrees per second. Thereupon, a reasonable maximum rate of turn relies upon around 7.2 degrees per second once the airspeed considered is close to 150kts [66]. Hence, although it can be adjusted in future works considering the detailed performance description of specific eVTOL vehicles models, we assume a maximum rate of turn of 7.2 degrees per second.

---

[17]In order to be considered safe, a trajectory must not present a conflict.

Moreover, the speed represents how long the aircraft flies per hour and, consequently, per second. Then, the position of the aircraft is a tuple (x,y) that represents a point in the Cartesian plane. In UAM operations, cruise segments are expected to be conducted in an altitude range close to 1000ft Above Ground Level (AGL) [39]. However, this requirement does not exclude vehicles crossing considering different altitudes. In this research, we consider early UAM operations, which requires special safety rules for ensuring the capability of this initiative once successful operations in the first years are vital. Thereupon, this research considers that the longitudinal separation requirement for each aircraft (i.e., 0.25NM for piloted and 0.5NM otherwise) is applied regardless of the altitude for ensuring a sparsity. This approach may be valuable in several situations (e.g., emergencies)

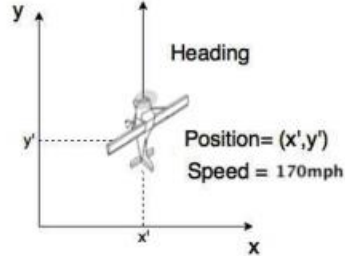

Figure 7: Example of an eVTOL vehicle state.

Source: The author.

Ticks drive the simulations conducted herein [18], which represents 1 second. For instance, consider that 1 knot (kt)[19] is equivalent to 1 Nautical Mile (NM)[20] per hour (h), that 1kt is equivalent to 0.000278NM per second (s), and that 170mph is approximately equal to 150 kts. An eVTOL operating with a speed of 150 kts (or 150NM/3600s), in 1 second (1 tick or 1 loop execution) it flies 0.0417 NM (150 × 0.000278).

Although each eVTOL model has its own performance specs and configurations, a reasonable en-route eVTOL airspeed range varies from 150mph to 200mph at 1000ft Above Ground Level (AGL) [9] [40]. Instead of simulating precisely the kinematics of the eVTOL vehicles, our goal in the simulation process is to provide an operational profile that is compatible to the eVTOL requirements [39], i.e., the main idea is to simulate a feasible trajectory profile that can be followed by UAM-enabled vehicles. Similarly to the simulation conducted [9], this research considers a constant en-route eVTOL airspeed of 170mph. This enables (i) the vehicles to follow the trajectories with a higher certainty factor, (ii) the system to be more predictable once arrival estimations become simpler, and (iii) easier adaptability and flexibility of the vehicles in terms of speed.

Nevertheless, given a specific trajectory point to be achieved, the eVTOL vehicle changes its heading for flying in the right direction. This process of changing the heading is illustrated in Figure 8. This change must respect the aforementioned maximum rate of turn, which is considered to be 7.2 degrees per second [66]. Finally, the eVTOL vehicles that reach the final point of its trajectory (i.e., its destination) exit the cruise airspace once the landing procedure is started. As the capacity of skyports is considered to be considerably large, multiple aircraft can be delivered to the same position.

Moreover, the eVTOL class, presents a set of attributes and methods. The attributes considered are:

· id: identifies the eVTOL vehicle with a unique number. This number is used in different moments for identifying specific vehicles;

· x: represents the position of the eVTOL vehicle in the x-axis, which can also be seen as a representation of the longitude;

· y: represents the position of the eVTOL vehicle in the y-axis, which can also be seen as a representation of the latitude;

· z: represents the position of the eVTOL vehicle in the z-axis, which can also be seen as a representation of the altitude;

---

[18]One tick represents one step in the simulation process and a time interval of one second. This means that, in one tick, the distance traveled by aircraft is the same distance traveled in one second.

[19]$1kt = 1.852 km/h = 0.868 kts$

[20]$1nm = 1.852 km = 0.868 NM$

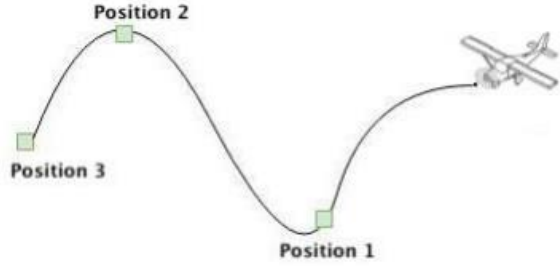

Figure 8: Changes applied to heading in order to reach the objective points.

Source: The author.

· target_x: represents the position in the x-axis of the final target of the eVTOL vehicle;

· target_y: represents the position in the y-axis of the final target of the eVTOL vehicle;

· target_z: represents the position in the z-axis of the final target of the eVTOL vehicle;

· hdg: represents the current heading of the eVTOL vehicle. This attribute varies from 0 up to 359.9;

· speed: represents the current speed of the eVTOL vehicle. This attribute varies from 130 (kts) up to 170 (kts);

· rate_of_climb: represents the maximum climbing rate of the particular vehicle;

· rate_of_descent: represents the maximum descending rate of the particular vehicle;

· type: identifies the type of the vehicle. The possible vehicle types are (i) piloted, (ii) remotely piloted, and (iii) self-piloted;

· hsep: represents the minimum horizontal separation requirement of the vehicle. In this research, minimum separations of 0.25NM for piloted vehicles and 0.5NM otherwise are considered;

· vsep: represents the minimum vertical separation requirement of the vehicle. In this research, the minimum vertical separation considered is 200ft;

· timestamp: this attribute highlights the scheduled time of the eVTOL vehicle regarding the simulation process, i.e., the time the vehicle starts its operation;

· objectiveList: represents the set of positions the eVTOL vehicle must follow, including the final position. This vector represents the trajectory of the aircraft considering all the assumptions;

· delivered: this attribute is a flag that identifies if the aircraft was delivered or not (i.e., if its operation was completed successfully);

· conflictList: this attribute represents a list of conflicting eVTOL vehicles according to the minimum horizontal separation. A safe trajectory leads the vehicle, from origin to destination, with an empty conflictList.

Furthermore, the methods considered in this class are:

· step: This method is responsible for conducting the eVTOL vehicle throughout the simulation process. As the UAM object calls this method, one execution of this method represents one second of operation in the simulation process as well. Furthermore, as 1 second is considered at a speed of 150kts according to the assumptions considered, the vehicle flies 0.0417NM in each execution of this method (as 1kt is equivalent to 0.000278NM per second, $150 \times 0.000278 = 0.0417$NM);

· add_obj_point: This method adds a new position to the vehicle's trajectory, i.e., a new position (x,y) is added to the objectiveList attribute;

· follow: This method is responsible for making sure the eVTOL vehicle is flying toward the next position of the trajectory. In order to accomplish this, a set of other methods are also called (e.g., move, calcAngleToPosition, and adjustHeading);

· move: This method changes the current position of the vehicle to the next position (i.e., the new position after one second) based on the current vehicle heading. In order to change the aircraft position, the applyMovementBasedOnHeading method is called. Finally, this method also checks the final position has been achieved;

· applyMovementBasedOnHeading: This method uses the distance the aircraft must move (i.e., 0.0417NM) and applies it accordingly to the current aircraft heading. In order to accomplish this, the cosine of the current heading is multiplied by the distance to move and, then, the result is added to the current position in the x-axis. Similarly, the sine of the current heading is multiplied by the distance to move and, then, the result is added to the current position in the y-axis. In this case, we use the cosine of the heading to get the x-axis movement and the Sine for the y-axis movement [76]. This process is illustrated in Listing 1.

```
1  def applyMovementBasedOnHeading(self, distanceToMove):
2          factorX = math.cos(math.radians(self.hdg))
3          factorY = math.sin(math.radians(self.hdg))
4          self.x += factorX*distanceToMove
5          self.y += factorY*distanceToMove
```

Listing 1: Implementation of the applyMovementBasedOnHeading method.

· calcAngleToPosition: This method is used for calculating the appropriate eVTOL vehicle heading for achieving the next objective point. The solution employed herein is based on the computation of the *atan* of the slope *m* [77];

· adjustHeading: This method verifies the difference between the current heading and the most appropriate heading for the particular situation. This enables the eVTOL vehicle to change its direction accordingly to the rate of turn assumed in this research (a maximum variation of 7.2 degrees per second);

Finally, Figure 4 highlights the levels each aircraft should fly depending on the heading adopted to achieve their respective destinations. Furthermore, Figure 9 depicts the entire operation of the eVTOL vehicles in this simulation tools. Firstly, the vehicle follows a vertical take-off procedure up to the take-off fix[21].

Then, the aircraft goes through its cruise trajectory. In this phase, the trajectory may lead the aircraft to change its flight level. After that, the aircraft flies to the landing fix, which is designed to prepare the eVTOL vehicles for the spiral landing[22]. Finally, the aircraft starts the spiral landing procedure in order to achieve the skyport.

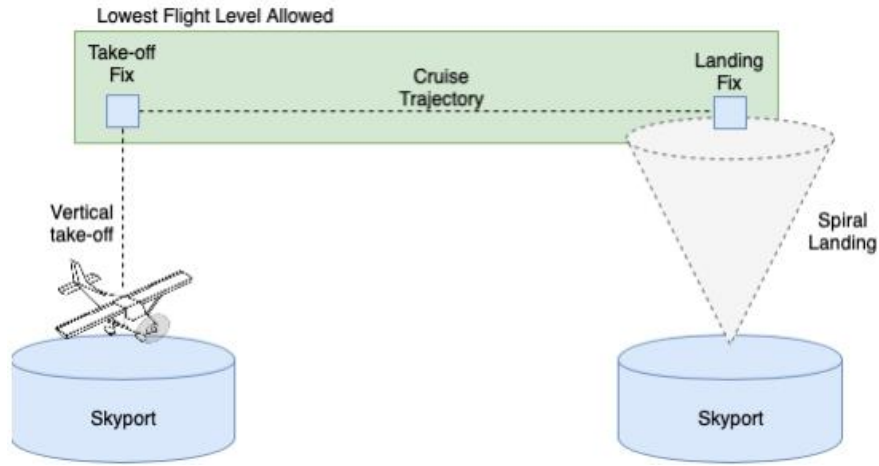

Figure 9: Example of an eVTOL vehicle state.

Source: The author.

---

[21]This position is placed at the same position of the skyport (x, y) and the altitude of the lowest flight level allowed. In this sense, the levels of the take-off fix are 1000ft and 1200ft (according to Figure 4), depending on the flight direction.

[22]The spiral landing is adopted once (i) decreasing the eVTOL vehicle altitude in a continuous landing procedure can present impacts on lower airspace areas (e.g., areas in which drones operate) and (ii) vertical landing procedures require a speed reduction of the eVTOL vehicle during cruise operations (which would require more elaborated landing procedures. As the main focus of this research relies on trajectories, we consider more detailed landing and takeoff procedures to be in the scope of future works.)

### 4.5.3 VideoMaker

The main goal of this class is to build a video for describing the trajectories of eVTOL vehicles. The process of building visual outputs enables a quicker understanding of the aspects of the solutions proposed. Furthermore, this is a static method, i.e., there is not an instance of such class, and the methods are directly accessed.

The videos created are composed of a group of frames. Each frame is a scatter plot created using Seaborn [78]. For instance, if one eVTOL vehicle is flying from a specific origin to a specific destination, each second of its movement (stored into the movements_df attribute of the UAM_manager object) is converted to scatter plot using the x- and y-axis as the longitudinal orientation.

The only method this static class presents is the build_video_df. The inputs considered are the history of the UAM operation, the plot size (x and y), and the full file name of the video to be created. Finally, it stores the video into the location pointed out.

A screenshot of a video produced by the VideoMaker class is illustrated in Figure 10. This scenario is illustrated in a 5NM×5NM region. The arrows highlight the meaning of each symbol during the simulation process. A blue circle identifies the eVTOL vehicle. Each eVTOL vehicle has a specific horizontal separation requirement depending on its type, and the color of the minimum separation also changes according to its size (0.25 or 0.5). Furthermore, some information about the operation is also presented: its type and id, the current speed (kts) and the current altitude (ft). Hence, the next objective point of each aircraft is also illustrated in the process as a blue circle. Finally, the written labels that bring more information to the process can also be hidden if it is necessary to improve the visualization.

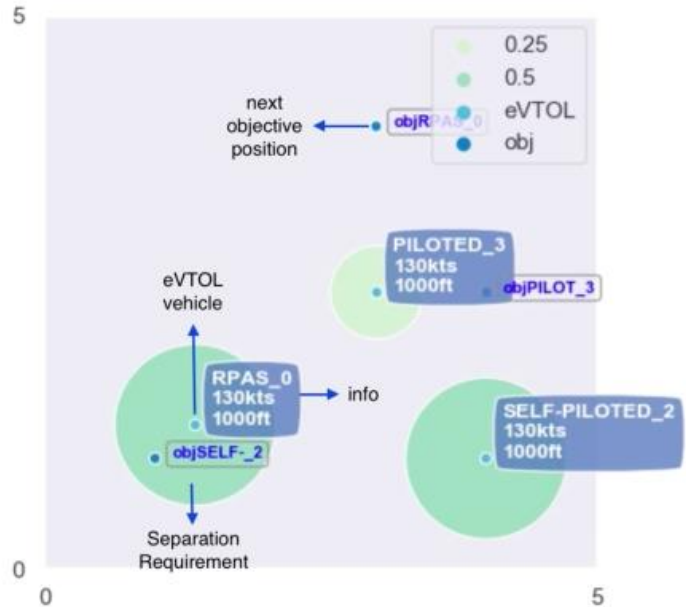

Figure 10: Screenshot of the Trajectory-Based UAM Operations Simulator (TUS).

Source: The author.

## 5 Experiments

In this Section, we present the simulation experiments for demonstrating the capabilities of the Trajectory-Base UAM Operations Simulator (TUS). This Section is organized as follows: Firstly, an experiment that aims at showing the execution of a simple UAM scenario is presented. Finally, an experiment that considers more complex maneuvers is presented. Each experiment presents the code used for building its respective scenario.

## 6 Experiment I

This experiment aims at demonstrating the capabilities of the TUS of simulation a specific scenario and to provide a visual representation of the flights conducted. As the aircraft included into the UAM are assumed to have safe

en

en

trajectories planned, our goal in this experiment is only to conduct the eVTOL vehicles throughout previously planned trajectories. Indeed, these trajectories were manually developed. The script for generating such scenario is highlighted in Listing 2.

```
1    from UAM import UAM
2    from eVTOL import eVTOL
3
4    uam = UAM()
5    ev1 = eVTOL(ev_id = 1, ev_type = 'rpas', origin = (10,5), destination = (20,5))
6    ev2 = eVTOL(ev_id = 2, ev_type = 'uas', origin = (10,15), destination = (20,15))
7    ev3 = eVTOL(ev_id = 3, ev_type = 'piloted', origin = (10,25), destination = (20,25))
8    ev4 = eVTOL(ev_id = 4, ev_type = 'piloted',origin = (20,10), destination = (10,10))
9    ev5 = eVTOL(ev_id = 5, ev_type = 'piloted',origin = (20,20), destination = (10,20))
10   uam.add_ev_list([ev1,ev2,ev3,ev4,ev5])
11   flight_duration = uam.simulate()
```

Listing 2: Script for building the scenario of experiment I.

Figure 11 presents the simulation execution and the operation of the aircraft[23]. The red dashed lines represent the trajectories defined for each eVTOL vehicles (in this case, all trajectories lead the vehicles to direct flights). The airspace operation time for delivering all the aircraft (considering take-off, cruise, and landing) was 1084s[24].

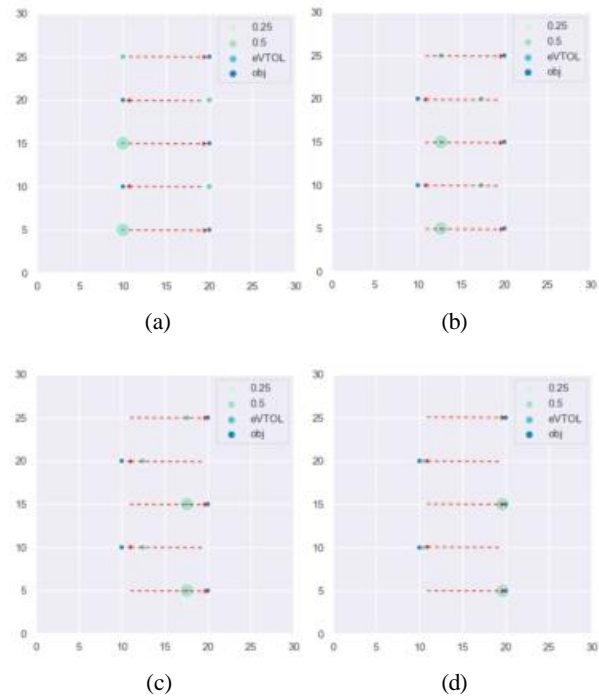

(a)      (b)

(c)      (d)

Figure 11: Simulation of eVTOL vehicles operation considering the trajectories defined.

Source: The author.

---

[23] In order to make the images and the simulation representation more precise, the labels were removed.

[24] This is not the processing time. It is time it takes to deliver all aircraft following the proposed trajectories (in this case, all aircraft are following a direct flight procedure).

## 7 Experiment II

The main goal of this experiment is to show the applicability of TUS in a situation that a specific eVTOL vehicle needs to be assigned to an alternative trajectory, which is different from the direct flight.

In this experiment, the scenario considered refers to a smaller area composed of a square of 5NM × 5NM. The script used for building this scenario is illustrated in Listing 3. Finally, both vehicles considered are going in the same direction in a direct flight approach.

```
1  from UAM import UAM
2  from eVTOL import eVTOL
3
4  uam = UAM()
5  ev1 = eVTOL(ev_id = 1, ev_type = 'rpas',origin = (4,4),destination = (2,2))
6  ev2 = eVTOL(ev_id = 2, ev_type = 'uas',origin = (4,2),destination = (2,4))
7  uam.add_ev_list([ev1, ev2])
8  flight_duration = uam.simulate()
```

Listing 3: Script for building the scenario of experiment II.

Figure 12 highlights the operation of those vehicles following direct flights. The results showed that these trajectories generate a conflict, i.e., this is not considered a safe solution. Indeed, one of these vehicles need an alternative trajectory in order to be delivered safely.

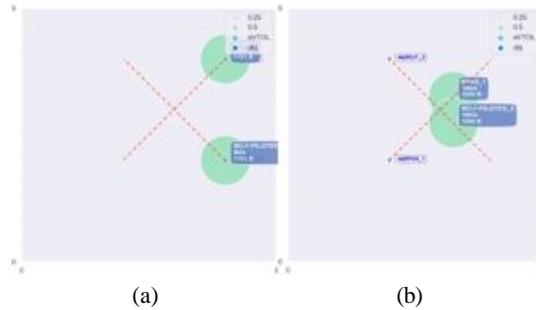

(a)           (b)

Figure 12: Simulation of eVTOL vehicles operation in experiment II considering the pre-defined trajectories.

Source: The author.

Moreover, this experiment aims to simulate a conflict detection in a specific situation. Once these trajectories are not considered safe, TUS stops its execution when the conflict is detected. One possible solution for this problem is to assign one of these eVTOL vehicles to an alternative trajectory. For example, if the eVTOL vehicle 1 is assigned to a trajectory composed of one additional point, a feasible solution can be found. Figure 13 illustrates a feasible solution in which the eVTOL vehicle 1 is assigned to a trajectory composed o one additional point ([4, 2, 1200]).

In this solution, eVTOL vehicles 1 and 2 were delivered in 691s and 399s, respectively. Furthermore, TUS did not detect any conflict once the vertical and horizontal separation standards were respected.

## 8 Conclusion

This research presented a simulation framework for measuring safety and effectiveness of Trajectory-Based UAM operations in urban environments considering the presence of both manned and unmanned eVTOL vehicles. This framework is intended to enhance trajectory planning capabilities by providing a suitable manner of measuring the effectiveness of each trajectory.

The architecture of the system was described as well as all assumptions adopted. After that, some experiments were conducted to show the applicability of our proposal in measuring different scenarios that can be faced in UAM operations.

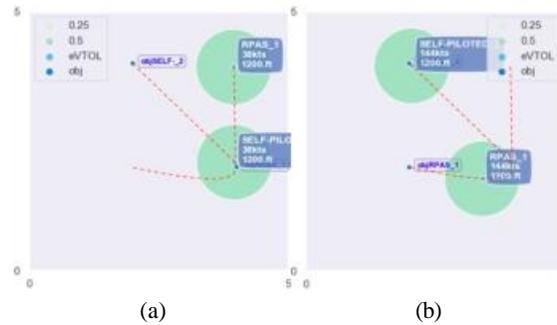

(a)          (b)

Figure 13: Simulation of eVTOL vehicles operation in experiment II considering an alternative trajectory for eVTOL vehicle 1.

Source: The author.

Finally, our intention with this contribution is to provide the research community with an approach for solving trajectory planning problems faced in different situations. Indeed, this simulation tool can be used for measuring the capabilities of different planning methods.

As future works, many areas can be explored. For example, more specific eVTOL models descriptions (e.g., speed and consumption), power consumption of eVTOL vehicles and their impact on trajectory planning, relationship between UAM operations and NAS operations, and interoperability of UAM and UTM.